\documentstyle[onecolumn,psfig,amsmath,times]{mn}

\title[General Relativistic Electromagnetic Fields of a
	Slowly Rotating Magnetized Neutron Star.]
	{General Relativistic Electromagnetic Fields of a
	Slowly Rotating Magnetized Neutron
	Star. II. Solution of the Induction Equations }

\author[Olindo Zanotti and Luciano Rezzolla]
	{Olindo Zanotti$^{(1)}$ and Luciano Rezzolla$^{(1),(2)}$	\\
								\\
	$^{(1)}$SISSA, International School for Advanced Studies,
	Via Beirut, 2-4 34014 Trieste, Italy			\\
	$^{(2)}$INFN, Department of Physics, University of
	Trieste, Via A. Valerio, 2 34127 Trieste, Italy 	\\
	} 

\pagerange{\pageref{firstpage}--\pageref{lastpage}}
\pubyear{2001}

\begin{document}

\maketitle
\label{firstpage}

\begin{abstract}
We have solved numerically the general relativistic
induction equations in the interior background spacetime
of a slowly rotating magnetized neutron star. The
analytic form of these equations was discussed in a
recent paper (Rezzolla {\it et al} 2001a), where
corrections due both to the spacetime curvature and to
the dragging of reference frames were shown to be
present. Through a number of calculations we have
investigated the evolution of the magnetic field with
different rates of stellar rotation, different
inclination angles between the magnetic moment and the
rotation axis, as well as different values of the
electrical conductivity. All of these calculations have
been performed for a constant temperature relativistic
polytropic star and make use of a consistent solution of
the initial value problem which avoids the use of
artificial analytic functions.  Our results show that
there exist general relativistic effects introduced by
the rotation of the spacetime which tend to {\it
decrease} the decay rate of the magnetic field. The
rotation-induced corrections are however generally hidden
by the high electrical conductivity of the neutron star
matter and when realistic values for the electrical
conductivity are considered, these corrections become
negligible even for the fastest known pulsar.
\end{abstract}

\begin{keywords}
relativity -- stars: neutron -- rotation -- magnetic
fields
\end{keywords} 

\date{Accepted 0000 00 00.
      Received 0000 00 00.}

\section{Introduction}
\label{intro}

	Irrespective of the origin of magnetic fields in
neutron stars, whether produced by thermoelectric effects
active in a thin layer below the star surface when the
temperature is much above $10^6 \rm{K}$ (see, for
instance, Wiebicke \& Geppert, 1996), or by a dynamo
action during the earliest stages of the convective
motions (see Thompson \& Duncan, 1993), or by post
core-collapse accretion of fall-back material after a
supernova explosion giving rise to a neutron star, a
secular decay of the magnetic field is expected as a
result of the finite electrical conductivity of the
stellar matter. The theoretical research in this area is
intense, pushed on by the observational evidence that
magnetic fields in neutron stars are decreasing with
increasing spin-down age. There is now a general
consensus about the possibility of improving the present
knowledge of the internal structure of neutron stars by
using the constraints from observations of the magnetic
field decay. This justifies the effort of taking into
account all of the possible factors which are supposed to
play a role during the decay of the magnetic field.

	Particularly interesting within this context are
the general relativistic corrections induced by the
presence of a strongly curved background spacetime. These
corrections have been investigated by a number of authors
(Ginzburg \& Ozernoy 1964, Anderson \& Cohen 1970,
Petterson 1974, Gupta {\it et al} 1998, Konno \& Kojima
2000) and with a number of different approaches some of
which are more rigorous (Geppert {\it et al}, 2000) than
others (Sengupta 1995, 1997). In recent related works,
Rezzolla {\it et al} (2001a, 2001b) have performed a
detailed analysis of Maxwell's equations in the external
and internal background spacetime of a rotating
magnetized conductor. As a result of this analysis, it
was possible to show that in the case of finite
electrical conductivity, general relativistic corrections
due both to the spacetime curvature and to the dragging
of reference frames are present in the induction
equations. Moreover, when the stellar rotation is taken
into account, each component of the magnetic field is
governed by its own evolutionary law, thus removing the
degeneracy encountered in the case of nonrotating
spacetimes. The purpose of this paper, which is the
natural extension of the work in Rezzolla {\it et al}
(2001a, hereafter paper I), is to quantify the general
relativistic effects related to rotation on the evolution
of the magnetic field. We have therefore solved
numerically the general relativistic induction equations
derived in Paper I for a relativistic polytropic star
with different values of the rotation period and of the
electrical conductivity. Each of the several calculations
performed here benefits from the consistent solution of
the initial value problem for a magnetic field which is
initially permeating a perfectly conducting relativistic
star. This approach avoids the use of artificial initial
data and provides a more accurate solution of the
induction equations.

	Overall, our results show that the rotation of
the star and of the background spacetime introduce a {\it
decrease} in the decay rate of the magnetic field. In
general, however, the rotation-induced corrections are
hidden by the high electrical conductivity of the neutron
star matter and are effectively negligible even for the
fastest known pulsar. Also in the absence of rotation,
the spacetime curvature introduces modifications to the
evolution of the magnetic field when compared with the
corresponding evolution in a flat spacetime. These
modifications depend sensitively on both the metric
functions of the interior spacetime and on the radial
profile of the electrical conductivity.  In the case the
star is modeled as a polytrope and the electrical
conductivity is assumed to be uniform in space and time,
the spacetime curvature generally increases the decay
rate of the magnetic field as compared to the flat
spacetime case, with this increase being dependent on the
compactness of the star.

	The paper is organized as follows: in Section 2
we discuss our treatment of the internal structure of the
star in the limit of slow rotation. Section 3 is devoted
to the solution of the induction equations derived in
paper I, with some emphasis on the numerical aspects and
in particular on the initial value problem. We show our
results in Section 4, whereas Section 5 contains the
conclusions. Throughout, we use a space-like signature
$(-,+,+,+)$ and a system of units in which $G = c =
M_{\odot} = 1$ (However, for those expressions of
astrophysical interest, we have written the speed of
light explicitly.). Partial spatial derivatives are
denoted with a comma.

\section{Stellar Structure}
\label{star}

	The background metric of a stationary,
slowly-rotating star at first order in the angular
velocity $\Omega$, is given by
\begin{equation}
\label{metric}
ds^2 = -e^{2\Phi(r)} dt^2 + e^{2\Lambda(r)}dr^2
	-2\omega(r)r^2\sin^2{\theta} dt d\phi + r^2
	\sin^2{\theta}d\phi^2
	 , \
\end{equation}
where $\omega(r)$ is the angular velocity of a
free-falling inertial frame. For realistic values of the
stellar magnetic field (i.e. $B=10^{11}-10^{13}$ G) we
can neglect the contribution of the electromagnetic
fields to the background spacetime geometry and determine
the internal structure of the star and its interior
spacetime after solving the following system of ordinary
differential equations (henceforth TOV system, from
Tolmann, 1939; Oppenheimer \& Volkoff, 1939)
\begin{eqnarray}
\label{TOV1}
\frac{dp}{dr}&=&-\frac{(p+e)(m + 4\pi
	r^3 p)}{r^2(1-2m/r)}  \ ,
\\ \nonumber
\label{TOV2}
\frac{dm}{dr}&=&4\pi r^2e \ ,
\\ \nonumber
\label{TOV3}
\frac{d\Phi}{dr}&=&\frac{m+4\pi
	 r^3 p}{r^2(1-2m/r)}= 
	-\frac{1}{e}\frac{dp}{dr}\left(1+\frac{p}{e}\right)^{-1}
	\ ,
\end{eqnarray}  
where $p(r)$ is the pressure, $e(r)$ is the energy
density and $m(r)$ is the mass enclosed within $r$. Once
an equation of state has been chosen, the TOV system can
be solved numerically together with the differential
equation for the Lense-Thirring angular velocity
$\omega(r)$ in the internal region of the star
\begin{equation}
\label{omega}
\frac{1}{r^3}\frac{d}{dr}\left[r^4
	e^{-(\Phi+\Lambda)}\frac{d\bar{\omega}}{dr}\right] +4\frac{d(
	e^{-(\Phi+\Lambda)})}{dr}\bar{\omega}=0 \ ,
\end{equation}
where $\bar{\omega}\equiv\Omega-\omega$. After selecting
a value for the central rest-mass density, the set of
differential equations (\ref{TOV1})\,--\,(\ref{omega}) is
solved from the centre of the star until the pressure
vanishes, thus determining the radius $R$. For the
integration of eq.~(\ref{omega}), the solution near the
centre of the star is simplified if we use the analytic
power series expansion $\bar{\omega}/\bar{\omega}_c\simeq
1 + 8\pi(e_c + p_c)r^2/5$, valid for $r\rightarrow 0 $
and where the label ``$c$'' refers to a quantity at the
centre of the star (Miller, 1977). Since in the vacuum
region of spacetime external to the star $\omega(r)=2
J/r^3$, with $J$ being the total angular momentum, we can
determine the two unknown quantities $J$ and $\omega_c$
by imposing continuity of the angular velocity and of its
first derivative at the surface.

	The interior of the star influences the magnetic
evolution either macroscopically, by affecting the metric
quantities which enter the induction equations, or
microscopically, through the electrical conductivity
$\sigma$ which, in turn, depends on the star's
temperature and chemical composition (see Urpin \&
Konenkov, 1997; Page {\it et al} 2000). Our attention is
here mainly focussed on assessing the contribution coming
from rotational effects in general relativity on the
decay of the magnetic field~\footnote{It should be
mentioned that general relativistic corrections can
appear also in the constitutive relations of the Maxwell
equations, such as in the general relativistic form of
Ohm's law (Ahmedov 1999). These corrections are usually
negligible in the electrodynamics of relativistic stars
and will be neglected here.}.  As a consequence, we will
neglect the thermal and rotational evolution of the
neutron star and simply consider a constant in time and
uniform in space electrical conductivity. This is an
approximation, but a necessary one to disentangle the
many different effects that intervene in the general
relativistic evolution of the magnetic
field. Furthermore, as will be discussed in Section 4,
the assumption of a uniform electrical conductivity does
not affect the role of a rotating background spacetime in
the evolution of the magnetic field.

	We model our relativistic stars as 
polytropes with equation of state
\begin{equation}
p = K \rho^{1+{1}/{N}} \ ,	
\end{equation}
where $\rho$, $K$, $N$ are the rest-mass density, the
polytropic constant and the polytropic index,
respectively. As ``fiducial'' model of neutron star we
consider a polytrope with index $N=1$, polytropic
constant $K=100$, and central rest-mass density
$\rho_c=1.28\times 10^{-3}$. In this case, the radius $R$
and the total mass $M$ obtained through the solution of
the TOV system are respectively $R=14.15$ Km and $M=1.40\
M_{\odot}$, yielding a compactness ratio $\eta=0.29$. The
rotation period usually chosen for this model is
$P=10^{-3}$ s.

\section{Evolution of the internal magnetic field}
\label{evolution}

	As mentioned in the Introduction, the presence of
the stellar rotation lifts the degeneracy found in the
case of a nonrotating star (Geppert {\it et al} 2000) and
three distinct induction equations regulate the general
relativistic evolution of the magnetic field. In this
Section we discuss the solution of the induction
equations for each of the magnetic field components. The
main difficulties encountered in the numerical solution
are related to the definition of a consistent initial
value problem and to the complex nature of the partial
differential equations when a misalignment between the
rotation axis and the magnetic dipole moment is
present. In the following we discuss the strategies
adopted to handle these difficulties.

\subsection{The Relativistic Induction Equations}

	The induction equations for the magnetic field of
a slowly rotating relativistic star with finite
electrical conductivity have been derived in paper I and
we briefly recall them here for completeness. All the
measurements are performed in the orthonormal tetrad
frame of a ``zero angular momentum observer'' (ZAMO) and
we assume that the spatial components of the magnetic
field four-vector in this frame are solutions of the
Maxwell equations in the separable form
\begin{eqnarray}
\label{ansatz_1}
&& B^{\hat r}(r,\theta,\phi,\chi,t) = 
	F(r,t)\Psi_1(\theta,\phi,\chi,t)\ ,
\\\nonumber\\ 
\label{ansatz_2}
&& B^{\hat \theta}(r,\theta,\phi,\chi,t) = 
	G(r,t)\Psi_2(\theta,\phi,\chi,t)\ ,
\\\nonumber\\
\label{ansatz_3}
&& B^{\hat \phi}(r,\theta,\phi,\chi,t) = 
	H(r,t)\Psi_3(\phi,\chi,t)\ ,
\end{eqnarray}
where $F, G, H$ and $\Psi_1$, $\Psi_2$, $\Psi_3$ account
for the radial and angular dependences, respectively.
Here, $\chi$ is the inclination angle of the stellar
magnetic dipole moment relative to the rotation axis and
the time dependence in $F, G, H$ is here due to the fact that
we are not considering an infinite electrical
conductivity but are allowing the magnetic dipole moment
to vary in time.

	At first order in $\Omega$, the angular
eigenfunctions $\Psi_i$ are not affected by general
relativistic corrections and assume the flat
spacetime expressions
\begin{eqnarray}
\label{psi1}
&& \Psi_1 = \cos\chi \cos\theta +
		\sin\chi \sin\theta \cos\lambda(t) \ ,
\\\nonumber\\ 
\label{psi2}
&& \Psi_2 = \cos\chi \sin\theta
		- \sin\chi \cos\theta \cos\lambda(t) \ ,
\\\nonumber\\
\label{psi3}
&& \Psi_3 = \sin\chi \sin\lambda(t) \ ,
\end{eqnarray}
where $\lambda(t) \equiv \phi - {\Omega} t$ is the
instantaneous azimuthal position (see Fig.~1 of paper I).
Assuming that the contribution of electric currents are
negligible\footnote{Because the magnetic field decay is
studied on timescales that are much longer than the
electromagnetic wave crossing time, this is a very good
approximation.} the general relativistic evolution
equations for the radial eigenfunctions $F(r,t)$,
$G(r,t)$, $H(r,t)$ are
\begin{eqnarray}
\label{evolF}
\frac{\partial F}{\partial t}\Psi_1\sin\theta 
	&=& \frac{c^2 e^{-\Lambda}}{4\pi\sigma r^2}\Bigg\{
	\left[e^{\Phi}r\left(G-H\right)\right]_{,r}\sin\chi\cos\lambda
	-2\left[\left(e^{\Phi}r G\right)_{,r}
	+e^{\Phi+\Lambda}F\right]\Psi_1 \sin\theta
\nonumber\\	
	& & \hskip 1.0 cm 
	-\frac{1}{4\pi\sigma}\sin\chi\sin\lambda\biggl\{
	\left[\omega r (H-G)\right]_{,r}
	(1 - 2\sin^2\theta) + 
	2{\omega}e^{-\Phi}\left[\left(e^{\Phi}r H\right)_{,r}+
	e^{\Phi+\Lambda}F\right]
	\sin^2\theta 
\nonumber\\	
	& & \hskip 0.2 cm 
	- \Omega r\left(G-H\right)\Phi_{,r}
	\left(1-2\sin^2\theta\right)\biggr\}
	\Bigg\} \ ,
\end{eqnarray}
\begin{eqnarray}
\label{evolG}
\frac{\partial G}{\partial t}\Psi_2
       & =& \frac{c^2}{4\pi\sigma r}\Bigg\{
        \frac{e^{\Phi}\left(G-H\right)}{r\sin^2\theta}
        \cos\theta\sin\chi\left[\cos\lambda+
        \frac{{\omega}e^{-\Phi}}{4\pi\sigma}\sin\lambda\right]
        +e^{-\Lambda}\left[e^{-\Lambda}\left(e^{\Phi}r G\right)_{,r}
        +{e^{\Phi}}F\right]_{,r}\Psi_2
\nonumber\\
        & & \hskip 0.2 cm
        -\frac{e^{-\Lambda}}{4\pi\sigma}\cos\theta\sin\chi\sin\lambda
        \bigg\{
        e^{-\Lambda}\left[\omega r
        \left(G-H\right)\right]_{,r}
        +\omega\left[F+e^{-(\Lambda+\Phi)}
        \left(r e^{\Phi} H\right)_{,r}\right]
        + \Omega\left[\Phi_{,r}e^{-\Lambda}r\left(G-H\right)\right]
        \bigg\}_{,r}\Bigg\} \ ,
\nonumber\\ \\ \nonumber\\
\label{evolH}
\frac{\partial H}{\partial t}\sin\lambda &=&
	 \frac{c^2 e^{-\Lambda}}{4\pi\sigma r}
	\left\{\left[e^{-\Lambda}\left(e^{\Phi}r H\right)_{,r}
	 +e^{\Phi}F\right]\left[\sin\lambda -
	\frac{{\omega}e^{-\Phi}}{4\pi\sigma}
	\cos\lambda \right]\right\}_{, r}
	 +\frac{c^2e^\Phi \left(G-H\right)}{4\pi\sigma r^2
	\sin^2\theta}\left[\sin\lambda -
	\frac{{\omega}e^{-\Phi}}{4\pi\sigma}
	\cos\lambda \right]\ .
\end{eqnarray}
Together with the evolution equations
(\ref{evolF})\,--\,(\ref{evolH}), the scalar functions $F$,
$G$, and $H$ also satisfy the constraint condition of
zero-divergence for the magnetic field
\begin{equation}
\label{zerodiv}
\left[\left(r^2 F\right)_{, r} + 2 e^{\Lambda}r G\right] 
	\sin\theta\left(\cos\chi\cos\theta +
	\sin\chi\sin\theta\cos\lambda\right)+
	e^{\Lambda}r \left(H-G\right)\sin\chi\cos\lambda
	= 0 \ .
\end{equation}
A rapid look at equations
(\ref{evolF})\,--\,(\ref{evolH}) shows that in a rotating
spacetime the evolution of the poloidal and toroidal
components are correlated and that an initially purely
poloidal magnetic filed can gain a toroidal component
during its evolution and vice-versa. In the case of a
nonrotating star, on the other hand, the three induction
equations (\ref{evolF})\,--\,(\ref{evolH}) are not
independent and the magnetic field evolution is described
by a single scalar function: $F$ (see Geppert et
al. 2000).

\subsection{Strategy of the Numerical Solution}
\label{numeric}

	The numerical solution of equations
(\ref{evolF})\,--\,(\ref{evolH}) is simplified if done in
terms of the new quantities
\begin{eqnarray}
&&\widetilde F \equiv r^2 F \ , 	
\\\nonumber \\ 
&&\widetilde G \equiv e^{\Phi} r G \ ,
\\\nonumber \\ 
&&\widetilde H \equiv e^{\Phi} r H \ ,
\end{eqnarray}
which, when the inclination angle $\chi$ is nonzero and
the electrical conductivity is uniform,
allow us to rewrite eqs.~(\ref{evolF})\,--\,(\ref{evolH})
schematically as
\begin{eqnarray}
\label{evF}
&&\frac{\partial \widetilde F}{\partial t} = f_1
	\widetilde F_{,rr} + f_2 \widetilde F_{,r} + f_3
	\widetilde F + f_4 \widetilde H_{,r} + 
	f_5 \widetilde G + f_6 \widetilde G_{,r} \ , 
\\\nonumber \\ 
\label{evG}
&&\frac{\partial \widetilde G}{\partial t} = 
	g_1 \widetilde G_{,rr} + g_2
	\widetilde G_{,r} + g_3 \widetilde G + 
	g_4 \widetilde F_{,r} + g_5 \widetilde F +
	g_6 \widetilde H_{,r} + g_7 \widetilde H \ ,
\\\nonumber \\ 
\label{evH}
&&\frac{\partial \widetilde H}{\partial t} = 
	h_1 \widetilde H_{,rr} + h_2
	\widetilde H_{,r} + h_3 \widetilde H + 
	h_4 \widetilde F_{,r} + h_5 \widetilde F 
	+ h_6 \widetilde G \ .   
\end{eqnarray}

\noindent Explicit expressions for the set of
coefficients {$f_i,g_i,h_i$} can be found in Appendix A.
For $\chi \ne 0$, the coefficients $f_i, g_i, h_i$ have
terms which are time-dependent trigonometric functions of
$\Omega t$ and, as a result, each of the
eqs.~(\ref{evF})\,--\,(\ref{evH}) is not a simple
parabolic equation describing a pure diffusive
phenomenon. In addition to a secular Ohmic decay, in
fact, there will be a periodic modulation produced by the
rotation of the star. This is evident if we look, for
instance, at the coefficient $f_1$ in Appendix A and
which is given by the sum of two terms. The first one is
the constant ``diffusion'' coefficient responsible for
the decay on a secular timescale. The second term, on the
other hand, represents the correction due to the stellar
rotation. The periodic modulation is produced by the
trigonometric function $\tan{\lambda}$ and varies
therefore on the dynamical timescale set by the rotation
period of the star, $P$. The presence of these periodic
terms spoils the parabolic character and makes the set of
eqs.~(\ref{evF})\,--\,(\ref{evH}) a mixed
hyperbolic-parabolic one.

	Although the integration of
eqs.~(\ref{evF})\,--\,(\ref{evH}) is complicated in the
general case, we are here favoured by the fact that all
of the terms proportional to $\Omega$ or to $\omega$
(i.e. all of the terms directly related to the stellar
rotation) scale like $\sigma^{-2}$ and that the
electrical conductivity in realistic neutron stars is
very high, ranging in the interval $10^{21} - 10^{28}
\rm{s}^{-1}$. As a result, the star's rotation period is
about twenty orders of magnitude smaller than the secular
timescale and can be ignored in the numerical solution of
the equations. In practice then, we set all of the
periodic time-varying terms to be constant coefficients
and solve the set of eqs.~(\ref{evF})\,--\,(\ref{evH}) as
a purely parabolic system. In this way we can capture the
secular decay without having to pay attention to the high
frequency modulation. In Section 4, where we discuss the
results of the numerical integration of the induction
equations~(\ref{evF})\,--\,(\ref{evH}), we will also
comment on the validation of this procedure.

	Another important aspect of the numerical
solution is the use of the zero-divergence constrain
equation~(\ref{zerodiv}). We do not need, in fact, to
integrate in time all of the
eqs.~(\ref{evF})\,--\,(\ref{evH}) but can restrict the
evolution to two of them and obtain, at each timestep,
the remaining unknown radial eigenfunction from the
solution of the constraint equation
(\ref{zerodiv}). Adopting this strategy in the numerical
solution reduces the computational costs and, most
importantly, enforces a constrained solution at each
timestep.

	Having three induction equations, we can follow
the decay of each component of the magnetic field
separately. The physically relevant quantity is however
the modulus of the magnetic field which, in the locally
flat spacetime of the ZAMO observer, is simply given by
$|B|=[(B^{\hat r})^2 + (B^{\hat{\theta}})^2 +
(B^{\hat{\phi}})^2]^{1/2}$. The evolution of this
quantity, evaluated at the surface of the star, is the
one that will be discussed in the remainder of the paper.

\subsection{The Initial Value Problem}
\label{ivp}

	The consistent solution of the initial value
problem for the general relativistic decay of the
magnetic field in a rotating neutron star suffers from
two difficult aspects. The first one is that at present
the initial topology and location of the magnetic field
in neutron stars can be only argued on the basis of some
assumptions, so that the magnetic field can either
permeate the entire star, or be confined in a layer close
to the stellar surface. The first configuration is more
plausible if the magnetic field is the final product of a
dynamo action amplification (see Thompson \& Duncan,
1993), while the second field configuration is more
realistic in a scenario in which the magnetic field is
originated by thermoelectric effects (Urpin {\it et al},
1986; Wiebicke \& Geppert, 1996). We here focus our
attention mostly on the case of a magnetic field
permeating the entire star, but in Section 4 we also show
how the decay of the magnetic field depends on the depth
of penetration inside the star, when simplified
assumptions on the microphysics at the crust-core
boundary are made.

	The second difficult aspect of the initial value
problem concerns the definition of an initial
configuration which is also solution of the general
relativistic Maxwell equations. A possible approach to
this problem is the one proposed by Geppert {\it et al}
(2000) (but see also Sang and Chanmugan, 1987), who have
considered the initial magnetic field to be described by
Stoke functions that represent, in flat spacetime, a
class of exact solutions of the induction equation. In
this case, the radial eigenfunction $\widetilde F(r)$ at
the initial time can be obtained from
\begin{eqnarray}
\label{stoke}
\widetilde{F}(r,t) &=& B_0\left[\frac{\sin(\pi
	r/R)}{\pi^2 r/R} - \frac{\cos(\pi r/R)}{\pi
	}\right]e^{-\pi c^2 t/4\sigma R^2} , \ \hspace{2cm} 0\leq
r\leq R \ ,
\end{eqnarray} 
for $t=0$, where $B_0$ is the initial surface magnetic
field at the magnetic pole. Because eq.~(\ref{stoke}) is
not a solution of the general relativistic Maxwell
equations, one expects an initial error being introduced
in the solution of the induction equations, but also that
this error should disappear rapidly as the solution tends
to the one satisfying the Maxwell equations.

	To circumvent the problem of an inaccurate
solution during the initial stages of the evolution and
in order to calculate an initial magnetic field which is
solution of the relativistic Maxwell equations, we here
treat the initial magnetic field as the one permeating a
perfectly conducting medium. In this case, Rezzolla et
al. (2001a, 2001b) have shown that consistent radial
eigenfunctions can be obtained after solving the
following set of equations [see (71)\,--\,(73) of paper
I]
\begin{eqnarray}
\label{ir_1}
\widetilde F_{, r} + 2 e^{\Lambda-\Phi}\widetilde G = 0 
	\ , && 
\\ \nonumber  \\ 
\label{ir_2}
\widetilde H_{, r} + \frac{e^{\Phi+\Lambda}}{r^2} 
	\widetilde F = 0\ , &&
\\ \nonumber\\
\label{ir_3}
\widetilde H - \widetilde G = 0  \ . && 
\end{eqnarray}
In particular, combining eqs.~(\ref{ir_1}) and
(\ref{ir_2}), we obtain a second-order differential
equation for the unknown radial eigenfunction
$\widetilde{F}$
\begin{equation}
\label{in_prof}
\frac{d^2\widetilde{F}}{d r^2} + \frac{d}{d r}(\Phi -
	\Lambda)\frac{d\widetilde{F}}{d r} - 2 e^{2\Lambda}
	\frac{\widetilde{F}}{r^2}=0  \ . 
\end{equation}
%

\begin{figure}
\centerline{
\psfig{file=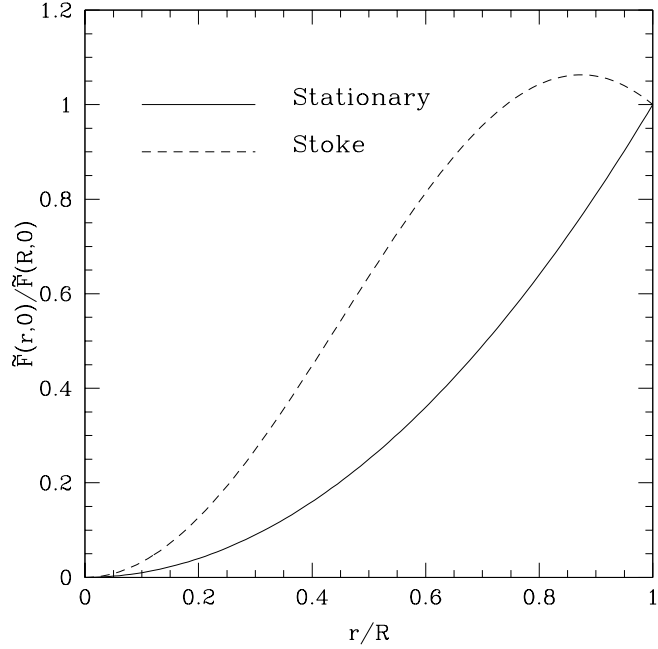,angle=0,height=9.0cm,width=9.0cm}
        }
\caption[]{\label{fig1} Possible initial values for the
radial eigenfunction $\widetilde{F}$ of the radial
component of the magnetic field normalized to the value
at the surface, shown as a function of the radial
position in the star. The solid line represents the
radial eigenfunction $\widetilde{F}$ as obtained from the
integration of the Maxwell eqs.~(\ref{in_prof}), while
the dashed line represents the Stoke profile
(\ref{stoke}). }
\end{figure}

	Equation (\ref{in_prof}) can be solved as a
two-point boundary value problem after specifying values
for the magnetic field at the edges of the numerical
grid. More specifically, the initial magnetic field at
the inner edge of the grid is chosen to be zero both when
the magnetic field permeates the whole star and when it
is confined to a crustal layer. On the other hand, the
initial magnetic field at the outer edge of the grid is
chosen to match a typical surface magnetic field for a
neutron star and is therefore set to be $B_0=10^{12}$
G. Once the initial profile for $\widetilde{F}$ has been
calculated through eq.~(\ref{in_prof}), the corresponding
initial values for $\widetilde{G}$ and $\widetilde{H}$
follow immediately from eqs.~(\ref{ir_1}) and
(\ref{ir_3}). As a comparison, we have also solved the
induction eqs.~(\ref{evF})\,--\,(\ref{evH}) using as
initial condition eq.~(\ref{stoke}) and the corresponding
eigenfunctions $\widetilde G$ and $\widetilde H$ again as
computed from the conditions (\ref{ir_1}) and
(\ref{ir_3}).

	Fig.~\ref{fig1} shows the initial values for the
two different prescriptions and, in particular, with a
solid line the initial profile as obtained through the
solution of the Maxwell eqs.~(\ref{in_prof}) and with a
dashed line the Stoke profile given by expression
(\ref{stoke}). The noticeable differences between the two
initial profiles provide a simple explanation of why the
use of Stoke's function produces an initially inaccurate
evolution (cf. Fig.~\ref{fig2}).

	The use of the strategy discussed above for the
calculation of the initial value problem clearly requires
the solution of an additional set of equations but it has
the advantage of removing the adjustment of the solution
during the initial stages of the decay and provides a
more accurate estimate of the magnetic field decay. A
discussion of this as well as a comparison with
evolutions performed with the Stoke function will be
discussed in Section 4. Finally, another aspect worth
stressing is that by using eq.~(\ref{in_prof}) we also
automatically satisfy appropriate boundary conditions at
the surface of the star.

\subsection{Boundary Conditions}

	In order to correctly solve the induction
equations (\ref{evF})\,--\,(\ref{evH}), it is essential
that appropriate boundary conditions are specified both
at the inner edge of the computational domain as well as
at the stellar surface.

	As for the initial value problem, the inner
boundary condition imposed during the evolution is that
of a zero magnetic field and is applied both when the
magnetic field permeates the whole star and when it is
confined to the crust. In the first case, this choice
guarantees a regular behaviour of the radial
eigenfuctions at the origin, while it reflects the
absence of magnetic field below the crust in the second
case. The evolution of the magnetic field has shown to be
quite sensitive to the boundary conditions imposed at the
stellar surface, but proper boundary conditions can be
derived if we assume that there are no electrical
currents on the surface and impose a matching between the
external and the internal solutions of the magnetic
field. The radial eigenfunctions $F(r),\ G(r)$, and
$H(r)$ outside the slowly rotating relativistic star have
been derived in paper I [see eqs.~(90)\,--\,(92) therein]
and are given by
\begin{eqnarray}
\label{f_of_r}
&&\widetilde{F}(r) = - \frac{3 r^2}{4M^3}
	\left[\ln N^2 + \frac{2M}{r}\left(1 +  \frac{M}{r}
	\right) \right] \mu
	\ ,
\\\nonumber\\ 
\label{g_of_r}
&& \widetilde{G}(r) = \frac{3 N^2}{4 M^2}
	\left[\frac{r}{M}\ln N^2 +\frac{1}{N^2}+ 1 \right] \mu 
	\ ,
\\\nonumber\\ 
\label{h_of_r}
&& \widetilde{H}(r) = \widetilde{G}(r) \ ,
\end{eqnarray}
where $N(r) \equiv (1 - 2M/r)^{1/2} = e^{\Phi}$ and $\mu$
is the magnetic dipole moment. Since the constraint
expressed by eq.~(\ref{ir_1}) holds also on the stellar
surface, we then have
\begin{equation}
\label{urca}
\widetilde {F}_{,r}\bigg\vert_R +
	2e^{\Lambda-\Phi}\widetilde G(R)=0 \ .
\end{equation}
Moreover, when electrical surface currents are not
present, we can use eq.~(\ref{f_of_r}) and
(\ref{g_of_r}) to express $\widetilde{G}(R)$ as
\begin{equation}
\label{g_of_f}
\widetilde{G}(R)=-\left(\frac{{\widetilde N}^2
 	M}{R^2}\right)\frac{R\ln {\widetilde N}^2/M
	+1/{\widetilde N}^2+ 1}{\ln {\widetilde N}^2 + 2M\left(1
	+ M/R \right)/R} \widetilde{F}(R) \ ,
\end{equation}
where ${\widetilde N} \equiv N(r=R)$. Straightforward
calculations allow to conclude that
\begin{equation}
\label{border2}
	R\widetilde {F}_{,r}\bigg\vert_R =
	\Pi(\eta)\widetilde F(t,R) \ ,
\end{equation} 
where $\Pi(\eta)$ is a constant given by
\begin{equation}
\Pi(\eta) \equiv  \frac{4\ln{(1-\eta)} + 2\eta
	(2-\eta)/(1-\eta)}{2\ln{(1-\eta)} +
	2\eta + \eta^2} \ ,
\end{equation}
with $\eta \equiv 2M/R$ being the compactness of the
star. The corresponding boundary conditions for
$\widetilde{G}$ and $\widetilde{H}$ are then easily
obtained by means of (\ref{h_of_r}) and (\ref{g_of_f}).

	Note that eq.~(\ref{border2}) coincides with the
boundary condition used by Geppert {\it et al} (2000) in
the case of a static, spherically symmetric background
geometry. This is due to the fact that, as discussed in
paper I, the corrections to the components of the
magnetic field enter at orders higher than the first one
in $\Omega$. Details on the numerical implementation of
the surface boundary conditions are presented in Appendix
B.

\section{Numerical Results}
\label{results}

	In order to integrate the set of induction
eqs.~(\ref{evF})\,--\,(\ref{evH}), we have built a
numerical code which implements the Crank-Nicholson
implicit evolution scheme and which provides second order
accuracy both in space and in time (see Morton \& Mayers,
1994). The accuracy of the code has been checked by
computing the time evolution of eq.~(\ref{stoke}) which
provides, in a flat spacetime, an exact solution of the
induction equation. The results obtained indicate that
the relative error between the numerical and the analytic
solutions over a timescale of three Newtonian Ohmic times
$\tau_{\rm ohm}\equiv 4\pi R^2 \sigma/c^2$, is always
below $0.5\%$ for the level of grid resolution usually
implemented in our calculations.

\begin{figure}
\centerline{
\psfig{file=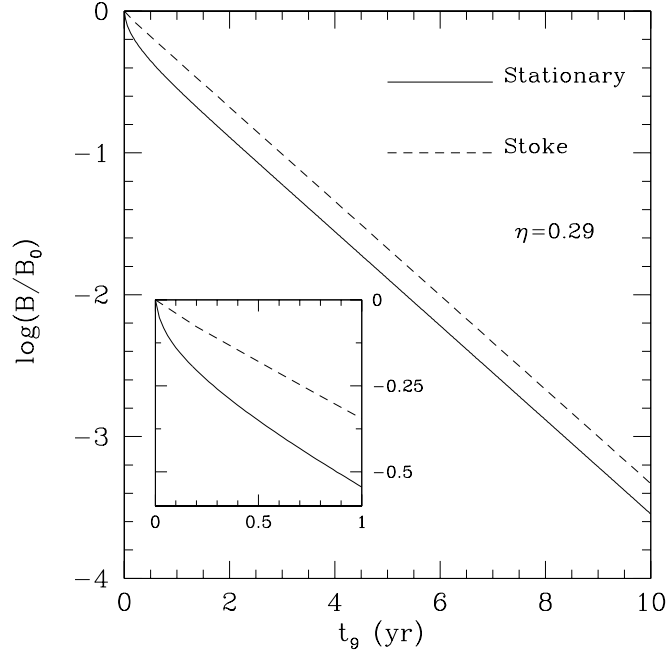,angle=0,height=9.0cm,width=9.0cm}
        }
\caption[]{\label{fig2} Difference in the decay of the
magnetic field when a consistent initial magnetic field
is used (solid line) or when a Stoke function is used as
initial condition (dashed line). The inset shows a
magnification of the evolution during the first $10^9$
yr. Here $\sigma=10^{25}s^{-1}$ and $P=10^{-3}s$.
See the main text for a complete discussion.}
\end{figure}

	Established the consistency and accuracy of the
code, we have proceeded to solve the general relativistic
induction equations for our relativistic rotating
star. As mentioned in Section 3.2, if the inclination
angle between the rotation axis and the dipolar magnetic
moment is nonzero, the secular decay has a periodic
modulation due to the stellar rotation. We have also
discussed that because the decay timescale and the
rotation period timescale differ for about twenty orders
of magnitude, we can neglect the time dependence (which
is $\propto \sin\lambda$) contained in each of the
coefficients $f_i$, $g_i$, $h_i$ and set the periodic
terms equal to an arbitrary constant value. To validate
this procedure and verify that the periodic modulation
does not affect the secular evolution, we have solved the
induction equations using different constant coefficients
and found that the secular results are indeed insensitive
to the value chosen for the constant coefficients. We
have also followed the solution of the complete set of
eqs.~(\ref{evF})\,--\,(\ref{evH}) (i.e. not considering
the time-periodic terms as constant) on a timescale which
is longer than the dynamical timescale but still much
smaller then the secular one. Also in this case we have
verified that the modulated evolution, which is
superimposed on the secular one, shows the small decrease
corresponding to the secular decay.

	Our discussion of the results starts by comparing
the evolution of eqs.~(\ref{evF})\,--\,(\ref{evH}) for
the two different prescriptions of the initial value
problem discussed in Sect.~\ref{numeric}
(cf. Fig.~\ref{fig1}) for our fiducial neutron
star. Before presenting the results of the comparison, it
is useful to discuss briefly the subtleties related to
the measure of the magnetic field time decay; as will
become apparent later, this is an important issue which
might lead to seemingly conflicting results. The gauge
freedom inherent in the theory of General Relativity
allows for the choice of arbitrary observers with respect
to which the measure of physically relevant quantities is
made. The choice of a certain class of observers might
rely on the mathematical advantages that this class may
have, but not all observers are physically suitable
observers. Locally inertial observers are certainly
preferable and in a rotating spacetime, as the one
considered here, ZAMO observers represent a natural
choice. Of course, there is is an infinite number of such
observers, each one performing his own measure of the
magnetic field decay, so that one should then select a
specific set of inertial observers on the basis of
physical considerations. The results presented in this
paper will be referred to a ZAMO observer on the surface
of the star and at a latitude $\theta=\pi/2$. The values
of the magnetic field measured by this observer and its
time evolution can then be converted to the equivalent
ones measured by other ZAMOs at different radial and
polar positions through simple transformations involving
the difference in the red-shifts and latitudes. Once the
choice of a suitable class of inertial observers is made,
it is also important that the results of the general
relativistic magnetic field decay are expressed using
appropriate units. In their work, Geppert {\it et al}
(2000) have quantified the decay of magnetic field in a
relativistic constant density, nonrotating star in terms
of the Newtonian Ohmic time. As we shall show below,
while this choice is acceptable for a constant density
star, it could be misleading in general.

	The two solutions of
eqs.~(\ref{evF})\,--\,(\ref{evH}) are presented in
Fig.~\ref{fig2} and show the decay of the magnetic field
, rescaled on a timescale $t_9 \equiv 10^9$ yr. It is
interesting to note that while the asymptotic decay rates
of the magnetic field are almost the same for the two
approaches, a final difference emerge. This is because
when using Stoke's function as initial data, the
evolution does not satisfy Maxwell's equations during the
initial stages (see the small inset in Fig.~\ref{fig2}),
but settles onto a constrained solution only after that
time. Moreover, the outer boundary conditions expressed
by eqs.~(\ref{g_of_f}) and (\ref{border2}) cannot be
satisfied exactly by Stoke's function and this introduces
an additional error. As a result, after a time $t \sim
t_9$ yr, the two solutions differ of about $45\%$, but
this difference does not grow further in time.

\begin{figure}
\centerline{
\psfig{file=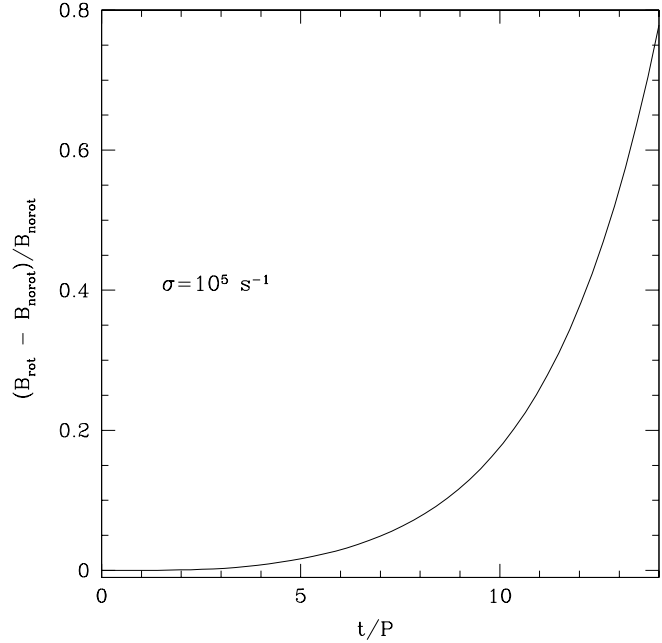,angle=0,height=9.0cm,width=9.0cm}
        }
\caption[]{\label{fig3} Relative difference in the
evolution of a magnetic field in a nonrotating star
$B_{\rm nonrot}$ and in a rapidly rotating one $B_{\rm
rot}$. The electrical conductivity is here set to be
$\sigma=10^5\ {\rm s}^{-1}$, while the star been set to
have a period $P=10^{-3}$ s.}
\end{figure}

	Next, we discuss the effects introduced by the
rotation of the star and of the spacetime. In this case
it is worth distinguishing the interest in finding a
general relativistic correction, from the impact that
these corrections actually have on the magnetic field
decay in a realistic rotating neutron star. As discussed
in Section 3.2, in fact, the high value of the electrical
conductivity in realistic neutron stars tends to make the
general relativistic corrections due to rotation rather
minute. In particular, we have found that when
considering an electrical conductivity $\sigma=10^{25}\
{\rm s}^{-1}$ in a rapidly rotating neutron star with one
millisecond rotation period, the relative difference in
the magnetic field after $15\ t_9$ yr is only one part in
$10^{12}$. Nevertheless, general relativistic,
rotation-induced corrections have an interest of their
own and these corrections can be more easily appreciated
if smaller (and therefore less realistic) values of the
electrical conductivity are considered. In
Fig.~\ref{fig3} we show the relative difference in the
evolution of a magnetic field in a nonrotating star,
$B_{\rm nonrot}$, and in a rapidly rotating one with a
millisecond period, $B_{\rm rot}$. In this case and just
for illustrative purposes, an electrical conductivity
$\sigma=10^5\ {\rm s}^{-1}$ has been considered. As can
be appreciated from the figure, the corrections due to
the rotation {\it decrease} the rate of decay of the
magnetic field and after a few rotation periods, the
fastly rotating star will maintain a magnetic field which
is about a factor of two larger than the one calculated
for the nonrotating star.  Overall, the results obtained
indicate that General Relativity does introduce, through
the rotation of the spacetime, new corrections to the
evolution of the magnetic field, slightly {\it
decreasing} its decay rate. This effect, however, is
usually hidden by the high electrical conductivity of the
stellar medium and can be neglected in general. The
results discussed above depend also on the inclination
between the rotation axis and the magnetic dipole moment,
with the decrease rate being larger for larger
inclination angles. In particular, for $\chi=\pi/2$, the
residual magnetic field after $10\ t_9$ yr is smaller of
a factor two as compared to the corresponding magnetic
field for an inclination $\chi=0$.

\begin{figure}
\centerline{ \hbox{
\psfig{file=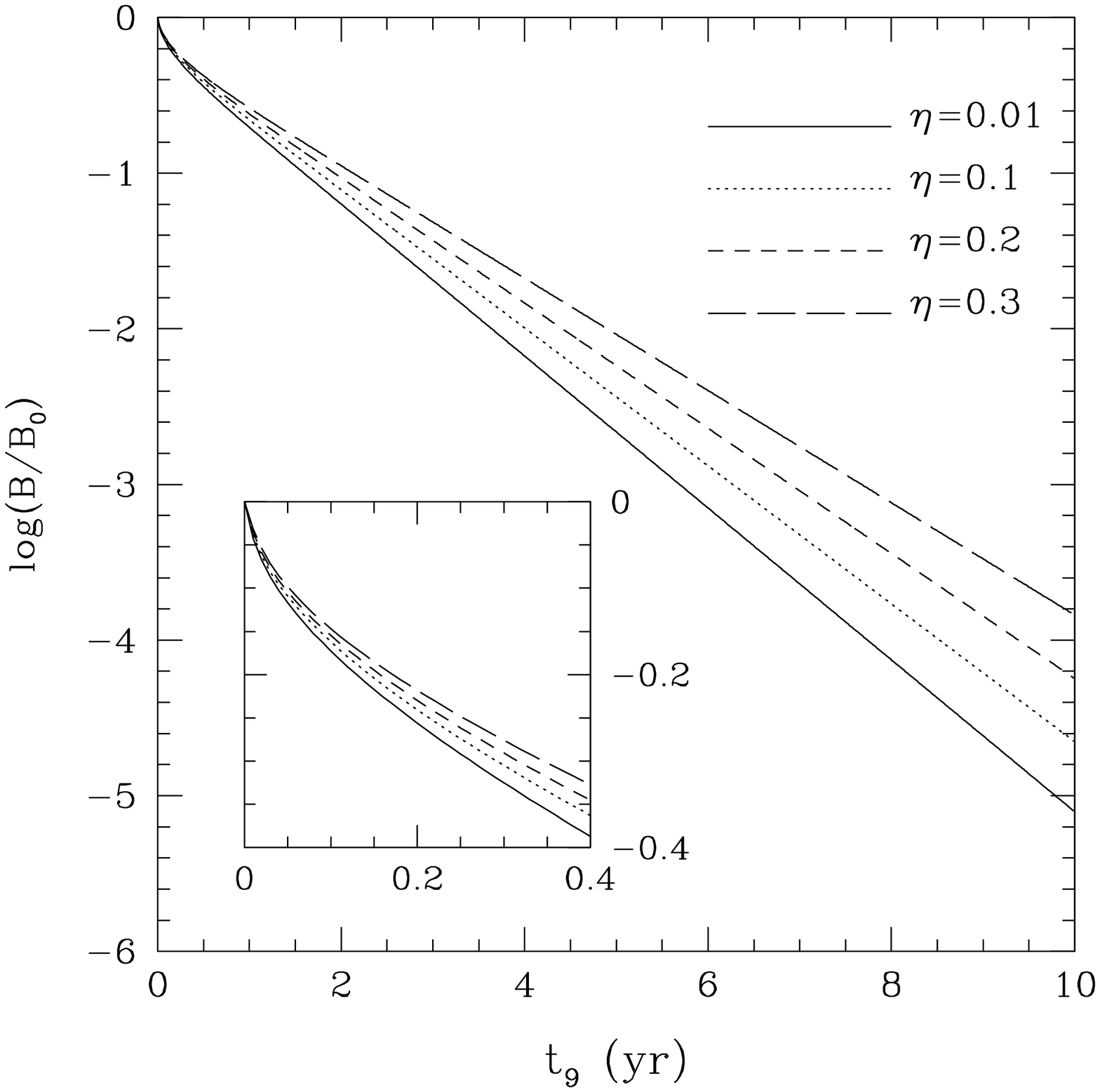,angle=0,width=8.5cm}
\hspace{0.125truecm}
\psfig{file=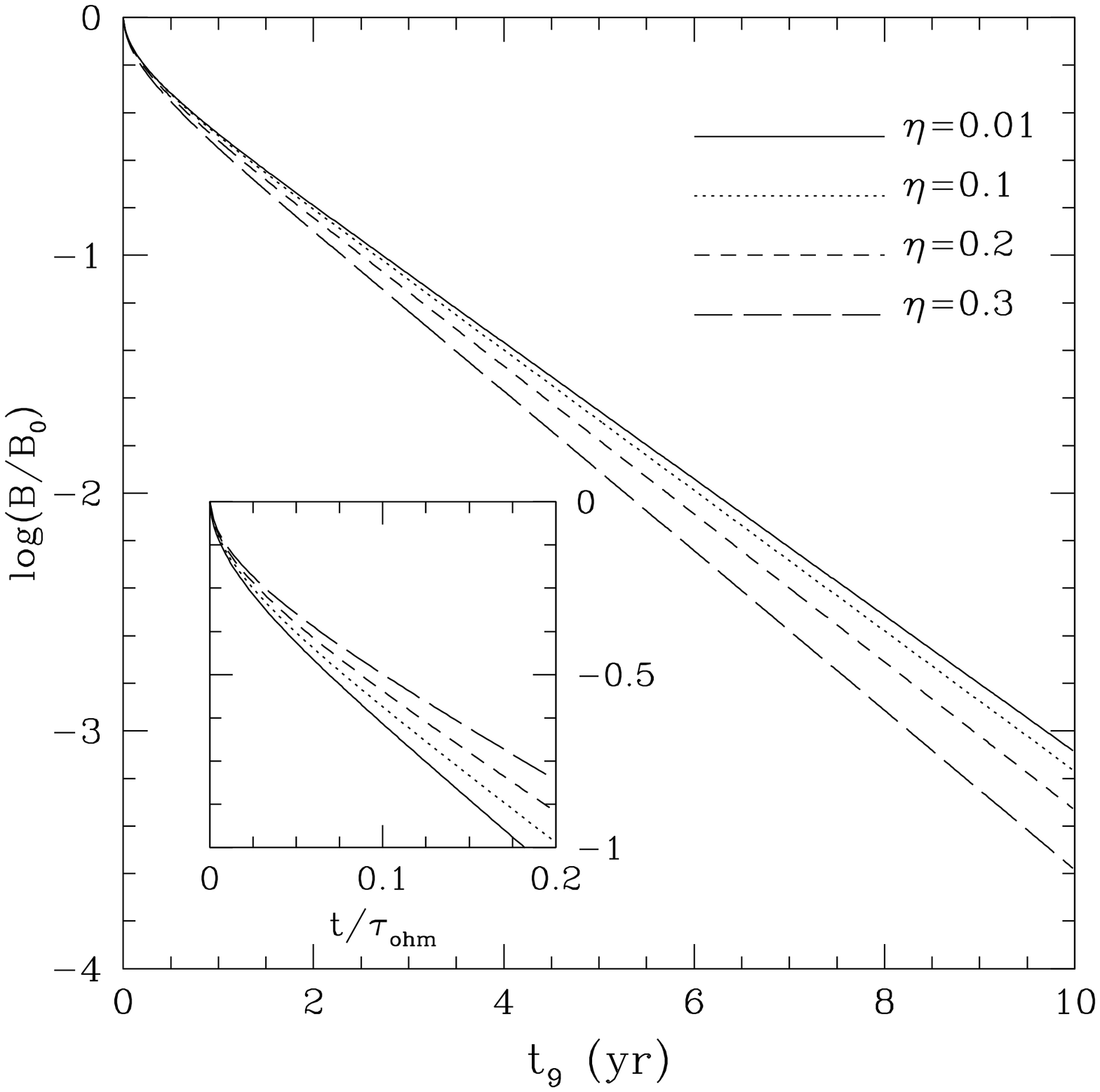,angle=0,width=8.5cm}}}
\caption[]{\label{fig4} Decay of the surface magnetic
field as measured by a ZAMO observer on the surface of
the star at a latitude $\theta=\pi/2$, expressed on a
timescale $t_9 \equiv 10^9$ yr. The left panel refers to
a constant density stellar model and shows an asymptotic
decay rate of the magnetic field which is {\it
decreasing} for increasing values of the stellar
compactness. The inset in the left panel focuses on the
initial stages of the evolution when the decay is
larger. The right panel, on the other hand, refers to an
$N=1$ polytropic stellar model and shows an asymptotic
decay rate which is {\it increasing} for increasing
values of the stellar compactness. Here the central
density is the free parameter determining the stellar
compactness. The small inset in the right panel of the
figure shows how the use of an Ohmic timescale as
normalizing unit can lead to erroneous interpretations.}
\end{figure}

	Next, we compare the results of our calculations
for a polytropic relativistic star with those for a
constant density star. This will provide a first
qualitative estimate of the importance of the metric
functions in the actual evolution of the magnetic
field. The results are presented in Fig.~\ref{fig4} with
the left panel referring to a constant density model and
the right one to our fiducial polytropic model.

	In the case of a constant-density star, we
confirm the results obtained by Geppert {\it et al}
(2000) and find that the evolution of the magnetic field
approaches an exponentially decaying behaviour, with an
asymptotic decay rate which is generally {\it decreasing}
with increasing stellar compactness. The inset in the
left panel of Fig.~\ref{fig4} shows in more detail the
initial stages of the magnetic field decay and allows to
appreciate that the magnetic field evolution is initially
following an exponential decay with decay rates which are
quite large but that then reach an asymptotic value after
about $10^8$ yr (Geppert {\it et al} 2000).

	In the case of a polytropic star, on the other
hand, the results in the right panel of Fig.~\ref{fig4}
show a behaviour which is the opposite to the one
encountered for a constant density model and that the
asymptotic decay rate of the magnetic field is {\it
increasing} with increasing stellar compactness. When a
uniform electrical conductivity is used, the explanation
behind the two distinct behaviours has to be found in the
deviations that emerge in the internal spacetime for the
two stellar models and in particular in the first radial
derivatives of the metric functions $\Phi$ and $\Lambda$
[cf. eqs. (\ref{TOV1})\,--\,(\ref{TOV3})]. These
deviations produce sensible quantitative differences in
the coefficients of eqs.~(\ref{evF})\,--\,(\ref{evH})
(see the Appendix for the explicit form of the
coefficients) which are then responsible for the increase
in the decay rate. It should also be remarked that the
behaviour shown in the left panel could be easily
reproduced, also in the case of a polytropic model, by
means of a suitably defined electrical conductivity. In
other words, the results presented in Fig.~\ref{fig4}
underline that a definitive conclusion on the general
relativistic evolution of the magnetic field cannot be
drawn until a more realistic treatment of the electrical
conductivity and of the equation of state is made.

	The inset in the left panel of Fig.~\ref{fig4}
can be used to explain the comment made above on the use
of relevant normalization units. In the inset, in fact,
we have plotted the same evolution shown in the main
panel but with the time being normalized in terms of the
Newtonian Ohmic time. Note that when we do so, the
overall behaviour is inverted and the decay rate of the
magnetic field in now decreasing for increasing stellar
compactness. This is clearly incorrect and the misleading
behaviour is due to the fact that the concept of an Ohmic
time is a purely Newtonian one and is therefore justified
only in a Newtonian context. A more suitable normalizing
unit for a nonrotating relativistic star would be the
general relativistic analogue of the Newtonian Ohmic
time: ${\widetilde \tau}_{\rm ohm}\equiv 4\pi R^2
e^{2\Lambda - \Phi}\sigma/c^2$ as can be derived from
eqs.~(\ref{evF})\,--\,(\ref{evH}) in the limit
$\Omega=0$. Using this normalization, we would recover
the correct behaviour, with a magnetic field asymptotic
decay rate generally {\it increasing} with stellar
compactness. Unfortunately the validity of ${\widetilde
\tau}_{\rm ohm}$ is limited to nonrotating stellar models
only. Because of the difficulties of defining an Ohmic
timescale for the induction equations of a relativistic
rotating star, we measure the magnetic field evolution
simply in terms of the time measured by our ZAMO
observer.

\begin{figure}
\centerline{
\psfig{file=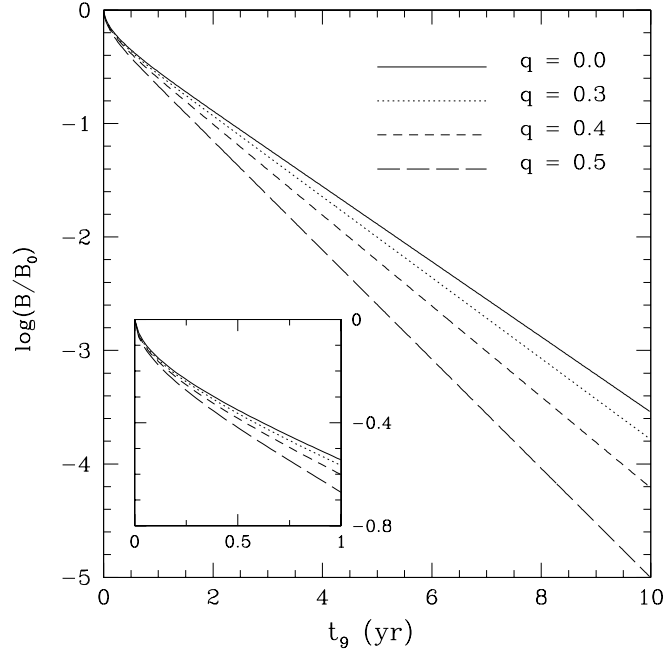,angle=0,height=9.0cm,width=9.0cm}
        }
\caption[]{\label{fig5} Decay of the surface magnetic
field when the magnetic field does not penetrate the
whole star. The different curves refer to different
values of the parameter $q\equiv R_{_{\rm IN}}/R$, with
$R_{_{\rm IN}}$ being the inner radius of the stellar
shell where the magnetic field is confined.}
\end{figure}

	Finally, we discuss the differences introduced in
the decay of the magnetic field when the latter is
confined to a spherical shell between an inner radius
$R_{_{\rm IN}}$ and the surface of the star. In this
case, the initial values for the radial eigenfunctions
are calculated self-consistently along the procedure
discussed in Section 3.3. In Fig.~\ref{fig5} we show the
evolution of the magnetic field in our fiducial neutron
star for different values of the parameter $q\equiv
R_{_{\rm IN}}/R$. Note that decreasing the volume in
which the magnetic field is confined has the effect of
{\it increasing} the decay rate of the magnetic field so
that if the initial magnetic field permeates about $90\%$
of the stellar volume ($q=0.5$), the residual surface
magnetic field after $10\ t_9$ yr is about a factor
thirty smaller than in the case the magnetic field
permeates the whole star ($q=0$). Although our analysis
does not take into account the microphysics of the
stellar interior and in particular the role played by the
chemical composition and by the temperature, it confirms
the Newtonian results of Urpin and \& Konenkov (1997) and
those of Page {\it et al} (2000), who have shown that the
magnetic field decay is slower for deeper magnetic field
penetration. Because this behaviour mimics the increase
in the decay rate produced by an increasing compactness
of the stellar model, it is essential to be able to
determine, prior to observations, the geometry and
location of the magnetic field within the neutron star
and to distinguish the different contributions to the
overall magnetic field decay.

\section{Conclusions}
\label{conc}

	In a recent paper, Rezzolla {\it et al}~(2001a)
have considered the general relativistic description of
the electromagnetic fields of a slowly rotating,
magnetized and misaligned neutron star. If the stellar
medium has a finite electrical conductivity it was shown
that the stellar rotation removes the degeneracy in the
evolution equations for the magnetic field and that three
distinct induction equations need to be solved to account
for the decay of the stellar magnetic field. In this
paper we have solved numerically the general relativistic
induction equations derived in paper I, investigating the
effects of different rotation rates, different
inclination angles between the magnetic moment and the
rotation axis, as well as different values of the
electrical conductivity. The aim of these numerical
calculations is that of quantifying the corrections
induced by general relativistic effects (both due to
spacetime curvature and to the stellar rotation) on the
evolution of the magnetic field of a slowly rotating
neutron star.

	In order to single out purely general
relativistic effects from those due to the microphysics
of the Ohmic dissipation, we have considered a simplified
physical description of the neutron star. In particular,
the star has been modelled as a polytrope rotating with a
fiducial period of one millisecond, the electrical
conductivity has been considered to be uniform inside the
star and we have not included a treatment to consider the
evolution of the stellar rotation and temperature (see
Page {\it et al} 2000). On the other hand, special
attention has been paid to a consistent solution of the
initial value problem and we have considered as initial
magnetic field the stationary solution of the general
relativistic Maxwell equations. In this way we have
avoided the use of initial magnetic field configurations
that are only approximate solutions of the Maxwell
equations (i.e. solutions of the Maxwell equations only
in the limit of flat spacetime).  Besides eliminating an
initial error during the initial stages of the magnetic
field decay, our prescription for the initial value
problem also provides a more accurate solution of the
Maxwell equations.

	The results of our computations have shown that
there exist general relativistic, rotation-induced
corrections to the evolution of the magnetic field. These
effects generally produce a {\it decrease} in the rate of
magnetic field decay. However, their contribution is
masked by the high value of the electrical conductivity
in realistic neutron stars and can be neglected in
general. Our calculations also indicate that general
relativistic effects not induced by the stellar rotation
can modify the time evolution of the magnetic field in a
magnetized star. Such effects are closely related to the
properties of the spacetime internal to the star and for
a polytropic stellar model with uniform electrical
conductivity these effects generally {\it increase} the
decay rate of the field. The validity of this conclusion
is however limited. Density gradients are in fact
expected in a realistic star and these will affect the
behaviour of the electrical conductivity which, in turn,
will influence the decay of the magnetic field.

	Our conclusions are that the general relativistic
evolution of the magnetic field in rotating neutron stars
can be studied with confidence already in a nonrotating
background spacetime. However, the role of a curved
background spacetime on the decay of the magnetic field
can be fully assessed only when the details of both a
realistic equation of state and of a realistic electrical
conductivity are carefully taken into account.  This will
be the subject of future work.

\section*{Acknowledgments}

We would like to thank Bobomurat Ahmedov and John Miller
for the numerous discussions. We have also appreciated
the useful comments of the referee, Ulrich
Geppert. Support for this research comes from the Italian
MURST and by the EU Programme ``Improving the Human
Research Potential and the Socio-Economic Knowledge
Base'' (Research Training Network Contract
HPRN-CT-2000-00137).

\appendix

\section[]{The numerical solution of the induction equations}

	In this Appendix we provide the explicit
expressions for the coefficients $f_i$, $g_i$, $h_i$
appearing in the new form of the induction equations
(\ref{evF})\,--\,(\ref{evH})
\begin{eqnarray}
 f_1 &=& \frac{c^2 e^{\Phi-2\Lambda}}{4\pi\sigma} -
\frac{c^2
e^{-2\Lambda}}{(4\pi\sigma)^2}\omega\tan\lambda(1-2\sin^2\theta)
\ ,\nonumber
\\
 f_2 &=& \frac{c^2
e^{\Phi-2\Lambda}}{4\pi\sigma}(\Phi_{,r} - \Lambda_{,r})
+ \frac{c^2
e^{-2\Lambda}}{(4\pi\sigma)^2}\tan\lambda(1-2\sin^2\theta)(\Omega\Phi_{,r}
+ \omega\Lambda_{,r} - \omega_{,r})\ ,\nonumber\\
 f_3 &=& \frac{-2 c^2 e^{\Phi}}{4\pi\sigma r^2} - \frac{2
\omega c^2}{(4\pi\sigma
r)^2}\sin\theta\frac{\Psi_3}{\Psi_1}\ , \nonumber\\
 f_4 &=& \frac{-2 c^2
e^{-\Phi-\Lambda}}{(4\pi\sigma)^2}\omega\sin\theta\frac{\Psi_3}{\Psi_1}\
, \nonumber\\
 f_5 &=& \frac{2 c^2
e^{-\Phi-\Lambda}}{(4\pi\sigma)^2}\tan\lambda(1-2\sin^2\theta)(\Phi_{,r}(\Omega
+ \omega) - \omega_{,r})\ , \nonumber\\
 f_6 &=& \frac{-2 c^2
e^{-\Phi-\Lambda}}{(4\pi\sigma)^2}\omega
\tan\lambda(1-2\sin^2\theta) \ ; \nonumber
\end{eqnarray}
\begin{eqnarray}
 g_1 &=& \frac{c^2 e^{\Phi-2\Lambda}}{4\pi\sigma} - \frac{c^2
e^{-2\Lambda}}{(4\pi\sigma)^2}\omega
\cos\theta\frac{\Psi_3}{\Psi_2}\ ,\nonumber\\
 g_2 &=& -\frac{c^2
e^{\Phi-2\Lambda}}{4\pi\sigma}\Lambda_{,r}+ \frac{c^2
e^{-2\Lambda}}{(4\pi\sigma)^2}\cos\theta\frac{\Psi_3}{\Psi_2}\left[\Phi_{,r}(2\omega-\Omega)
+ \omega\Lambda_{,r} -2\omega_{,r}\right]\ ,\nonumber\\
 g_3 &=& \frac{c^2
e^{\Phi}}{(4\pi\sigma)r^2}\frac{\cos\theta\sin\chi}{\sin^2\theta\Psi_2}(\cos\lambda
+ \frac{\omega e^{-\Phi}}{4\pi\sigma}\sin\lambda)\ ,  \nonumber\\
 &&-\frac{c^2
e^{-2\Lambda}}{(4\pi\sigma)^2}\cos\theta\frac{\Psi_3}{\Psi_2}\left[\omega_{,rr}-\omega_{,r}(\Lambda_{,r}+2\Phi_{,r})+\Phi_{,rr}(\Omega-\omega)+(\omega-\Omega)(\Phi_{,r}^2+\Phi_{,r}\Lambda_{,r})\right]\ ,\nonumber\\
g_4 &=& \frac{c^2
e^{\Phi-\Lambda}}{(4\pi\sigma)r^2}\left[e^{\Phi} -
\frac{\omega}{4\pi\sigma}\cos\theta\frac{\Psi_3}{\Psi_2}\right]\ ,\nonumber\\
g_5 &=&\frac{c^2
e^{2\Phi-\Lambda}}{(4\pi\sigma)r^2}(\Phi_{,r}-\frac{2}{r})-\frac{c^2
e^{\Phi-\Lambda}}{(4\pi\sigma
r)^2}\cos\theta\frac{\Psi_3}{\Psi_2}(\omega_{,r}-\frac{2\omega}{r})\ ,\nonumber\\
g_6 &=& -\frac{c^2
e^{-2\Lambda}}{(4\pi\sigma)^2}\cos\theta\frac{\Psi_3}{\Psi_2}\left[\Phi_{,r}(\omega-\Omega)-\omega_{,r}\right]\ ,\nonumber\\
g_7 &=&-\frac{c^2 e^{-2\Lambda}}{(4\pi\sigma)^2}\cos\theta\frac{\Psi_3}{\Psi_2}\left[\Phi_{,rr}(\omega-\Omega)-\omega_{,rr}+\Phi_{,r}^2(\Omega-\omega)+\Phi_{,r}\Lambda_{,r}(\Omega-\omega)+\omega_{,r}(2\Phi_{,r}+\Lambda_{,r})\right]\ ;\nonumber 
\end{eqnarray}

\begin{eqnarray}
h_1 &=& \frac{c^2
e^{\Phi-2\Lambda}}{4\pi\sigma}(1-\frac{\omega
e^{-\Phi}}{4\pi\sigma}\cot\lambda)\ , \nonumber\\
h_2 &=&\frac{c^2
e^{\Phi-2\Lambda}}{4\pi\sigma}\left[\frac{e^{-\Phi}}{4\pi\sigma}\cot\lambda
(\omega\Phi_{,r}+\omega\Lambda_{,r}-\omega_{,r}) -
\Lambda_{,r}\right]\ , \nonumber\\ 
h_3 &=& -\frac{c^2
e^{\Phi}}{(4\pi\sigma)r^2\sin^2\theta}\left(1-\frac{\omega
e^{-\Phi}}{4\pi\sigma}\cot\lambda\right)\nonumber\\
h_4 &=& \frac{c^2
e^{2\Phi-\Lambda}}{(4\pi\sigma)r^2}\left[1-\frac{\omega
e^{-\Phi}}{4\pi\sigma}\cot\lambda\right]\ , \nonumber\\
h_5 &=& \frac{c^2
e^{2\Phi-\Lambda}}{(4\pi\sigma)}\left[(\Phi_{,r}-\frac{2}{r})(1-\frac{\omega
e^{-\Phi}}{4\pi\sigma}\cot\lambda) +
\frac{e^{-\Phi}\cot\lambda}{4\pi\sigma}(\omega\Phi_{,r}-\omega_{,r})\right]\ , \nonumber\\
h_6 &=& \frac{c^2
e^{\Phi}}{(4\pi\sigma)r^2\sin^2\theta}\left(1-\frac{\omega
e^{-\Phi}}{4\pi\sigma}\cot\lambda\right)\ . \nonumber
\end{eqnarray}

\section[]{Implementing surface boundary conditions}

	This Appendix shows how the surface boundary
conditions expressed by equations (\ref{g_of_r}) and
(\ref{border2}) can be implemented in a numerical
code. By adopting the standard finite-difference notation
in which $u^n_j\equiv u(x_j,t^n)$ and assuming a uniform
radial grid with $J$ gridpoints, the finite difference
form of eq.~(\ref{border2}) is given by
\begin{equation}
\label{border3}
\widetilde F^n_{J+1} - \widetilde F^n_{J-1} = 2\Pi(\eta)\Delta x \widetilde
F^n_J/R , \
\end{equation}
where $\Delta x=x_j-x_{j-1}$, $\Delta t=t^{n+1}-t^n$, 
The unknown value of $\widetilde F^{n+1}_{J}$, comes after 
introduction of (\ref{border3}) into the
Crank-Nicholson scheme, centered at J; lengthy but
straigthforward calculation give
\begin{eqnarray}
\label{border4}
\widetilde F^{n+1}_{J} &=& \frac{\alpha(f_1^{n}\widetilde F^n_{J-1} +
	f_1^{n+1}\widetilde F^{n+1}_{J-1}) + \widetilde
	F^n_{J}\left[1+f_3^n \Delta t/2 - f_1^n\alpha +
	\Pi(\eta)\alpha \Delta x(f_2^n \Delta x/2 +
	f_1^n)/R\right]} \nonumber\\
\vspace{0.5cm}
&& +\frac{f_4^n \alpha \Delta x(\widetilde
	H^n_{J+1}-\widetilde H^n_{J-1})/2+f_5^n
	\widetilde G^n_{J}\Delta t 
	+f_6^n\alpha \Delta x(\widetilde
	G^n_{J+1}-\widetilde G^n_{J-1})/2}{1-f_3^{n+1}\Delta t/2 +
	f_1^{n+1}\alpha - \Pi(\eta)\alpha \Delta x(f_2^{n+1}/2 \Delta x +
	f_1^{n+1})/R} , \
\end{eqnarray} 
where $\alpha=\Delta t/\Delta x^2$. The are still two
unknowns entering (\ref{border4}), i.e. $\widetilde
G^n_{J+1}$ and $\widetilde H^n_{J+1}$. However, they
represent the external solution and by using
eqs.~(\ref{h_of_r}) and (\ref{g_of_f}), they can be
written as
\begin{eqnarray}
\widetilde G^n_{J+1}&=&\widetilde H^n_{J+1}=
	-\frac{\tilde{N}^2 M}{(R+\Delta x)^2}\frac{(R+\Delta x)\ln
	\tilde{N}^2/M + 1/\tilde{N}^2 +
	1}{\ln \tilde{N}^2 + 2M \left(1 +
	M/(R+\Delta x)\right)/(R+\Delta x)}\widetilde F^n_{J+1} ,\
\end{eqnarray} 
where $\tilde{N}$ is the value of $N$ at $R+\Delta x$
\begin{equation}
\tilde{N} \equiv N\bigg\vert_{R+\Delta x}=
	\left(1-\frac{2M}{R+\Delta x}\right)^{1/2} .\
\end{equation}
The updated values of $\widetilde G$ and $\widetilde H$
now follow immediately from (\ref{g_of_f}) with time
evolved value of $F_{J}^{n+1}$ given by (\ref{border4}).

\label{lastpage}

\begin{thebibliography}{99}

\bibitem{a99}{Ahmedov, B.~J., 1999. Phys. Lett. A, 256, 9.}
\bibitem{ac70}{Anderson, J.~L., Cohen, J.~M., 1970. Astrophys.
        Space Science, 9, 146.}
\bibitem{dt92}{Duncan, R. C., Thompson, C., 1992.
        ApJ, ~392, L9.}
\bibitem{gu94}{Geppert, U., Urpin V., 1994, MNRAS ,
        271, 490.}
\bibitem{gpz99}{Geppert, U., Page, D., Zannias, T., 2000.
	 Phys. Rev. D, 61, 123004}
\bibitem{go64}{Ginzburg, V. L., Ozernoy, L. M., 1964. Zh. Eksp. Teor.
        Fiz., 47, 1030.}
\bibitem{getal98}{Gupta, A., Mishra, A., Mishra, H.,
	Prasanna, A.~R., 1998, Class. Quantum Grav. 15, 3131.}
\bibitem{h67}{Hartle, J. B., 1967. ApJ, 150, 1005.}
\bibitem{ht68}{Hartle, J. B., Thorne, K. S., 1968.
	ApJ, 153, 807.}
\bibitem{kg01}{Konenkov, D., Geppert, U., 2001,
	MNRAS, 325, 426.}
\bibitem{kk00}{Konno K., Kojima Y., 2000, Prog. of
	Theor. Phys. 104, 1117.}
\bibitem{mm94}{Morton, K.~W., Mayers, D.~F., 1994, {\em Numerical
	Solution of Partial Differential Equations},
	Cambridge University Press}
\bibitem{m77}{Miller, J.~C.,
	1977. MNRAS 179, 483.}
\bibitem{ov39}{Oppenheimer, J.~R., Volkoff, G.M., 1939.
	Phys. Rev., 55, 374.}
\bibitem{pgz00}{Page, D., Geppert, U., Zannias, T., 2000, 
	Astron. Astrophys., 360, 1052.}
\bibitem{p74}{Petterson, J.~A., 1974. Phys. Rev., D10, 3166.}
\bibitem{pg97}{Prasanna, A.~R., Gupta, A., 1997, 
	Il Nuovo Cimento B, 112, 1089.}
\bibitem{rama}{Rezzolla, L., Ahmedov, B.~J., Miller,
	J.~C., 2001a, MNRAS, 322, 123}
\bibitem{ramb}{Rezzolla, L., Ahmedov, B.~J., Miller,
	J.~C., 2001b, Found. of Phys., 31, 1051.}
\bibitem{sal94}{Salgado, M., Bonazzola, S., Gourgoulhon,
	E., Haensel, P., 1994. Astron. Astrophys. 291, 155.}
\bibitem{sc87}{Sang Y. and Chanmugam G., 1987, ApJ, 323, L61.}
\bibitem{s95}{Sengupta, S., 1995. ApJ, 449, 224.}
\bibitem{s97}{Sengupta, S., 1997. ApJ, 479, L133.}
\bibitem{td93}{Thompson, C., Duncan, R., 1993., ApJ, 408, 194.}
\bibitem{t39}{Tolmann, R.C., 1939. Phys. Rev., 55, 364.}
\bibitem{uk97}{Urpin, V.,  Konenkov, D., 1997.
	 MNRAS, 292, 167.}
\bibitem{ws83}{Wasserman, I., Shapiro, S.~L., 1983.
	ApJ, 265, 1036.}
\bibitem{wg96}{Wiebicke, H. J., Geppert, U., 1996.
	Astron. Astrophys. , 309, 203.}

\end{thebibliography}
\end{document}